\documentstyle[sprocl,epsf,here,amsopn]{article}

\def\Journal#1#2#3#4{{#1} {\bf #2}, #3 (#4)}
\def\APNY{{\em Ann.\ Phys.\ (N.Y.)}}
\def\JMP{\em J. Mat.\ Phys.}
\def\RMP{\em Rev.\ Mod.\ Phys.}
\def\NPA{{\em Nucl.\ Phys.} A}

\def\NP{\em Nucl.\ Phys.}
\def\PLB{{\em Phys.\ Lett.} B}
\def\PRL{\em Phys.\ Rev.\ Lett.}
\def\PRA{{\em Phys.\ Rev.} A}
\def\PRC{{\em Phys.\ Rev.} C}
\def\ZPA{{\em Z. Phys.} A}
\def\be{\begin{equation}}
\def\ee{\end{equation}}
\def\bea{\begin{eqnarray}}
\def\eea{\end{eqnarray}}
\def\etal{{\em et al.}}
\newcommand{\Figurebb}[9]{
   \begin{figure}[H]\begin{center}
   \leavevmode
   \epsfysize=#7cm
   \epsfbox[#2 #3 #4 #5]{#6}
   \par
   \parbox{#8cm}{
   \caption[figure]{\renewcommand{\baselinestretch}{0.8}
   \small \hspace{-0.3truecm}#9}
   \label{#1}}
   \end{center}
   \end{figure}
}
\def\fig#1{Fig.~\ref{#1}}
\def\eq#1{(\ref{#1})}
\DeclareMathOperator{\sign}{\mathrm{sign}}
\def\bgrk#1{\mbox{{\boldmath $#1$ \unboldmath}}\!\!}
\def\hom{\hbar\omega}

\begin{document}

\title{SEMICLASSICAL CALCULATION\\
       OF SHELL EFFECTS IN DEFORMED NUCLEI}

\author{M. BRACK, CH. AMANN}

\address{Institute for Theoretical Physics, University of Regensburg\\
         D-93040 Regensburg, Germany}

\maketitle

\vspace*{-4.7cm}
\begin{flushright}
 {\bf TPR-00-14}
\end{flushright}
\vspace*{3.88cm}

\abstracts{We summarize recent work in which the shell effect, which
causes the onset of the mass asymmetry in nuclear fission, could be 
explained semiclassically in the framework of the periodic orbit 
theory. We also present new results for the inclusion of a 
spin-orbit interaction in the semiclassical calculation of the
level density.}

\section{Introduction}

The explanation of the asymmetry in the fragment mass distributions of
many actinide nuclei has long been a stumbleblock in
fission theory. Shortly after the advent of the shell model, Lise
Meitner\,\cite{lise} related the mass asymmetry to shell structure,
in particular to magic nucleon numbers in one of the fragments. Indeed,
since the classical liquid drop model (LDM) favours symmetric
shapes,\,\cite{cosw} the mass asymmetry had to be a quantum shell
effect. Johansson\,\cite{sjoh} attempted to calculate fission barriers
by summing the single-particle energies of a Nilsson-type shell model
including left-right asymmetric deformations. However, due to their
wrong average behaviour at large deformations, he could not obtain
finite barriers. Only after the introduction of the ingenious {\em
shell-correction method} by Strutinsky\,\cite{stru} this became
possible, launching large-scale investigations of the shell structure
in fission barriers.\,\cite{nils,fuhi,nix} The resulting picture of the
{\em double-humped fission barrier} was nicely confirmed in many
detailed experiments.\,\cite{exp}

One interesting outcome was that the {\em onset} of the left-right
asymmetry of the nuclear shapes starts quite early during the fission
process: in nuclei like $^{240}$Pu, already the outer fission barrier
is unstable against octupole-type deformations.\,\cite{asym} This
quantum shell effect could be related to a few specific diabatic
single-particle states which were particularly sensitive to the
asymmetric deformations.\,\cite{gumn} Today, we know that symmetric and
asymmetric fission modes may coexist in the same nucleus.\,\cite{bifi}
Of course, the mass distributions of the fission fragments can only be
predicted from a dynamical theory involving inertial
parameters.\,\cite{maru} But interestingly enough, the most probable
experimental mass ratios were found to be roughly equal to those of the
nascent fragments obtained statically at the asymmetric outer
barrier.\,\cite{fuhi,grei} The onset of the {\em mass asymmetry} in
nuclear fission was thus established as a {\em quantum shell effect}
due to specific single-particle states in the deformed average
potentials of the nucleons, which could not be understood in the
classical liquid drop model.

However, new developments have taken place over the last 25 years to
describe quantum shell effects {\em semiclassically} in the framework
of the {\em periodic orbit theory (POT).} Through the so-called {\em 
trace formula}, the oscillating part $\delta g\,(E)$ of the level 
density of a quantum system can be related to properties of the 
periodic orbits of the corresponding {\em classical system}. This 
approach was systematically developed by Gutzwiller;\,\cite{gutz} an 
alternative derivation valid for billiard systems (i.e., particles 
enclosed in a cavity with ideally reflecting walls) was given by Balian 
and Bloch.\,\cite{bb} Strutinsky \etal\,\cite{stma,does} generalized 
the Gutzwiller theory and used it for the semiclassical calculation of 
shell-correction energies in various shell-model potentials (neglecting, 
however, the spin-orbit interaction). In particular, they successfully 
explained the systematics of nuclear ground-state deformations in terms 
of the leading shortest periodic orbits.\,\cite{does} Similarly, the 
onset of the fission mass asymmetry could recently be explained
semiclassically.\,\cite{fis1} For a comprehensive presentation of the
POT, the trace formula and its further refinements, and its applications 
to finite fermion systems, we refer to a recent monograph.\,\cite{book}

In Sect.\ \ref{sec:asym} we summarize the recent work on the fission
asymmetry done by one of us (M.B.) in collaboration with P. Meier, S.
M. Reimann, and M. Sieber.\,\cite{fis1,fis2,fis3} In Sect.\
\ref{sec:so} we present some new semiclassical results for the level 
density of a three-dimensional deformed potential, including the 
spin-orbit interaction which so far has been omitted in applications 
of the POT to the calculation of shell structure in nuclear physics.

\section{Semiclassical explanation of the onset of mass asymmetry}
\label{sec:asym}

In our recent semiclassical investigation of the fission mass
asymmetry,\,\cite{fis1,fis2} we have approximated the nuclear mean
field by a cavity with constant volume and reflecting walls, neglecting
spin-orbit, Coulomb, and pairing interactions and considering only one 
kind of particles. We use the parameterization ($c,h,\alpha$) of
Ref.\,\cite{fuhi} to define the boundary of the cavity. Here $2c$ is
the length of the nucleus along the symmetry axis in units of the
radius $R_0$ of the spherical cavity (given by $c$=1, $h$=$\alpha$=0),
$h$ is the neck parameter, and $\alpha\neq 0$ yields left-right
asymmetric shapes. In $^{240}$Pu, e.g., the maximum of the symmetric
outer barrier is found\,\cite{fuhi} near $c=1.53$, $h=\alpha=0$. 

The semiclassical trace formula for the
shell-correction energy $\delta E$ reads\,\cite{stma,book}
\begin{equation}
\delta E \simeq \sum_{po}{\cal A}_{po}(E_F)
                \left(\frac{\hbar}{T_{po}}\right)^{\!2}
                \cos\left[\frac{1}{\hbar}S_{po}(E_F)
                                   -\sigma_{po}\frac{\pi}{2}\right].
\label{esc}
\end{equation}
The sum in \eq{esc} is taken over {\it all} periodic orbits ($po$), 
but the gross-shell effects are dominated by the shortest periodic
orbits of the system. 
$S_{po}=\oint_{po} {\bf p}\cdot d{\bf q}$ 

\Figurebb{barview}{110}{495}{514}{700}{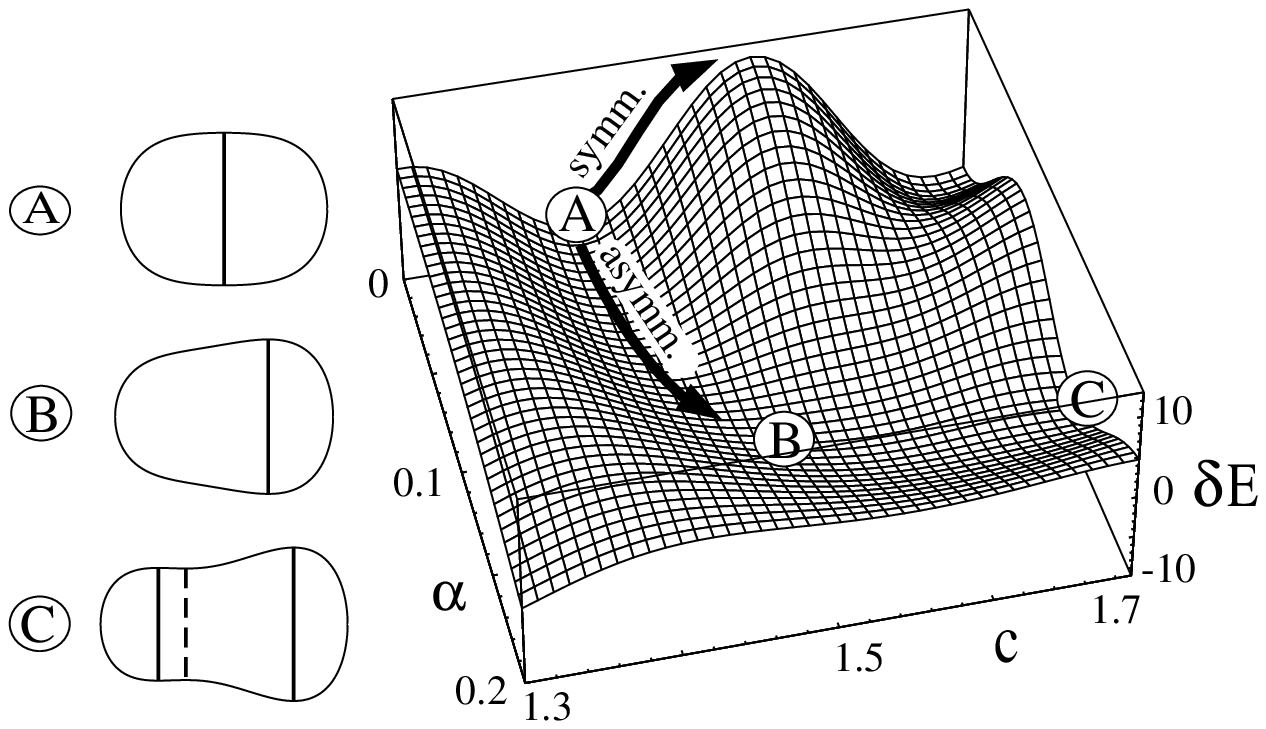}{5.1}{11.8}{
{\it Right}: Semiclassical deformation energy $\delta E(c,\alpha)$
($h$=0) in a perspective 3D plot (from Ref.\,\cite{fis1}). 
The arrows `symmetric' and `asymmetric' show two alternative 
fission paths. {\it Left}: shapes along the asymmetric path. Planes 
of the leading periodic orbits are shown by vertical lines (solid 
for stable and dashed for unstable orbits). 
}

\noindent
are the action integrals,
$T_{po}=dS_{po}/dE$ the periods, and $\sigma_{po}$ are phases related
to the number of conjugate (or focal) points along the periodic orbits. 
The amplitude ${\cal A}_{po}$ of an orbit depends on its stability and 
its degeneracy; it includes a factor that exponentially damps 
contributions from orbits with large periods $T_{po}$. All quantities 
in \eq{esc} are evaluated at the Fermi energy $E_F$.

In \fig{barview} we show a perspective 3D plot of the deformation
energy $\delta E(c,\alpha)$ ($h=0$) obtained from Eq.\ \eq{esc}. The
missing effect of the spin-orbit interaction was compensated by
adjusting the only parameter of the model, $E_F$, such that the 
isomer minimum is found at $c=1.42$, $h=\alpha=0$, corresponding to 
the quantum result for $^{240}$Pu in the realistic Strutinsky
calculations.\,\cite{fuhi} We see that the semiclassical model
correctly yields the instability of the outer barrier against the
asymmetry, and a realistic value ($\sim 0.13$) of the parameter
$\alpha$ at the asymmetric saddle point. Inclusion of the LDM energy,
which varies very little in the deformation region shown here, does
not change this result qualitatively. To the left in \fig{barview} we 
show the nuclear shapes at three points (A,B,C) along the adiabatic 
fission path. The vertical lines indicate the planes perpendicular to 
the symmetry axis in which the shortest periodic orbits are found. 
These are simply the polygons with $p$ reflections inscribed in the 
circular cross section of the cavity with these planes. A fast 
convergence of $\delta E$ with increasing $p$, i.e., with the length 
of the periodic orbits, was found in Ref.\,\cite{fis2} ($p=2$ and 3 
were essentially sufficient); other periodic orbits were also shown 
to be negligible.

We thus obtain a simple physical explanation of the onset of the mass
asymmetry of fission in terms of very few classical periodic orbits.
The valley of steepest descent through the deformation energy surface
is simply given\,\cite{fis1} by the requirement that the action of
the leading periodic orbits be stationary: $\delta S_{po}=0$. We have
also shown\,\cite{fis3} that the diabatic states which cause this
asymmetry effect quantum-mechanically have their probability maxima
exactly on the planes containing the leading periodic orbits, and that
a semiclassical quantization of the quasi-regular motion near these
planes reproduces the energy levels of the diabatic quantum states
surprisingly well.

Our semiclassical picture is only qualitative. It is not meant to 
replace the Strutinsky (or Hartree-Fock) type calculations, but to 
give an intuitive physical understanding of the mass asymmetry using
classical mechanics. Of course, one has to question if the neglect of 
the spin-orbit interaction can be justified since it is known to 
crucially affect the shell structure. Locally, its effect could here 
be simulated (like in Ref.\,\cite{does}) by adjusting the Fermi 
energy. But a semiclassical calculation including spin-orbit effects 
is certainly desirable.  

\section{Inclusion of spin-orbit interaction in POT}
\label{sec:so}

We present now some new results including a spin-orbit interaction
in the semiclassical trace formula. The cavity model of the previous 
section leads to problems with the standard (Thomas) form of the 
spin-orbit interaction which contains the gradient of the local 
potential. We therefore use for $V(\bf r)$ a 3-dimensional deformed
harmonic-oscillator potential and add a generalized spin-orbit term
which in the spherical limit is proportional to 
${\bf l}\cdot{\bgrk\sigma}$:
\begin{equation}
H=\frac{1}{2m}\,p^2+V({\bf r})
            +\kappa\,{\bf B}({\bf r,p})\cdot{\bgrk\sigma}
\label{defosc}
\end{equation} 
with 
\be
V({\bf r})= \frac{m}{2}\sum_{i=x,y,z} \omega_i^2 r_i^2\,, \qquad\quad 
{\bf B}({\bf r,p}) = {\bgrk \nabla}V({\bf r}) \times {\bf p}\,.
\label{Vho}
\ee
In Eq.\ \eq{defosc}, ${\bgrk\sigma}=(\sigma_x,\sigma_y,\sigma_z)$ are 
the Pauli matrices. We allow for different frequencies $\omega_i$ to 
mimic the shell-model potential of a deformed light nucleus. The 
quantum-mechanical eigenvalues of \eq{defosc} were obtained by an 
exact diagonalization in the basis of $V(\bf r)$ up to a maximum 
energy of $\sim 150\,\hom_x$. 

A first clue to the classical behaviour of this system is gained from 
a Fourier transform of the oscillating part $\delta g\,(E)$ of the 
level density from the energy to the time domain. To emphasize the 
gross-shell structure, the level density is Gaussian averaged over an 
energy range $\gamma<\,\hom_i$. Due to the energy scaling 
behaviour\,\cite{new} of the Hamiltonian \eq{defosc}, the peaks in the 
Fourier spectrum lie at the time periods $T_{po}$ of the classical 
orbits that dominate $\delta g\,(E)$. In \fig{ft} we show the Fourier 
spectra obtained for a case with {\it irrational} frequency ratios, 
both without and with spin-orbit interaction. For $\kappa=0$, we 
clearly see three 

\Figurebb{ft}{65}{503}{593}{700}{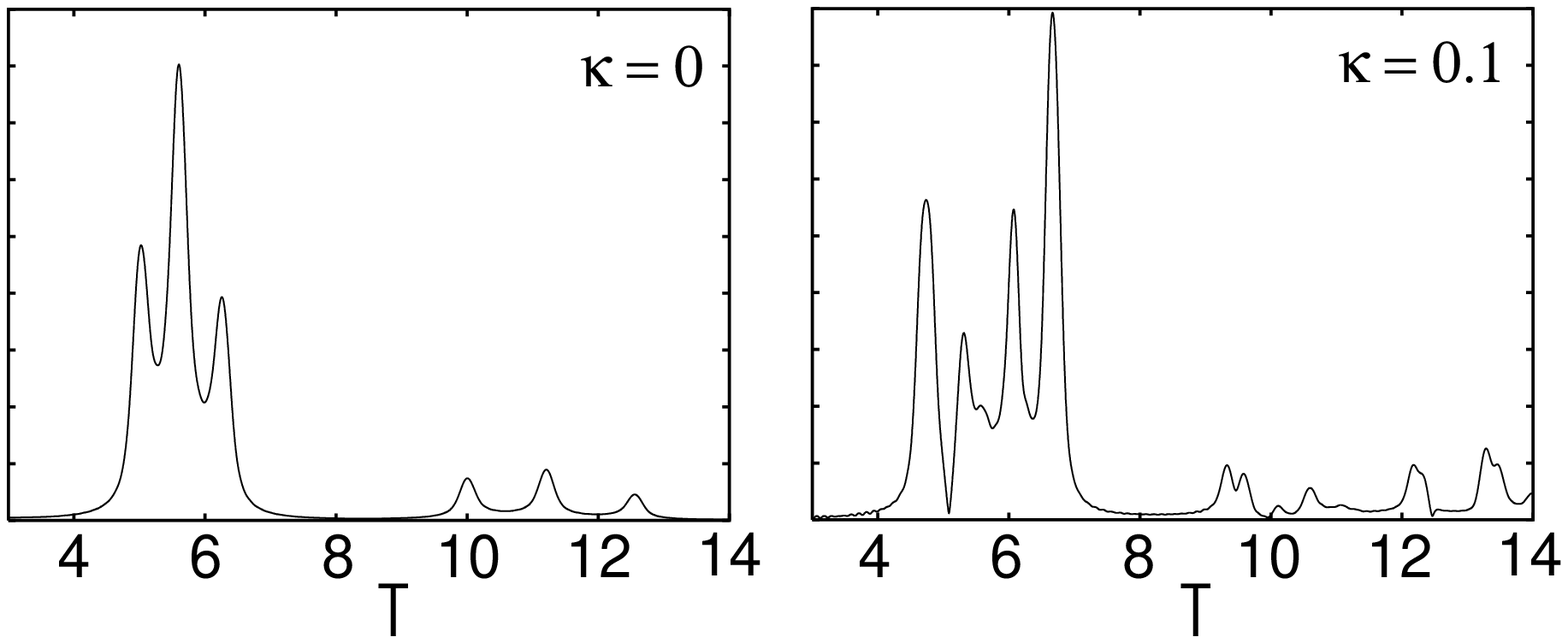}{3.1}{11.8}{
Squared Fourier amplitudes (in arbitrary units) of $\delta g\,(E)$
for the anisotropic harmonic oscillator \protect\eq{defosc} with
frequencies $\omega_x=1$, $\omega_y=1.12128$, $\omega_z=1.25727$, 
both without (left) and with spin-orbit interaction (right). Energy 
eigenvalues up to $\sim 50\,\hbar\omega_x$ were included; the 
Gaussian averaging range was $\gamma=0.2\,\hom_x$.
}

\noindent
main peaks which correspond exactly to the periods 
$T_i=2\pi/\omega_i$ of the only three periodic orbits of this system, 
which are the librations along the principal axes.\,\cite{book} At 
twice these periods, three small peaks corresponding to their second 
harmonics are visible. For $\kappa=0.1$, a more complicated peak 
structure is observed, which will be explained in the following.       

The difficulty in the semiclassical description is that the 
Pauli operators in \eq{defosc} act on a two-component spinor and 
couple its components in a non-trivial way. Littlejohn and 
Flynn\,\cite{lit1} have developed a theory whose basic idea is to diagonalize 
the Hamiltonian (2$\times$2) matrix locally in phase space after 
its Wigner transformation, and to expand both $H_W({\bf r,p})$ and 
the diagonalizing unitary matrix in powers of $\hbar$. Two problems 
arise hereby. First, on those surfaces in phase space where 
${\bf B}({\bf r,p})=0$, the eigenvalues become degenerate and the 
method breaks down (so-called {\it mode conversion}). Second, one 
obtains gauge-transformation dependencies of the diagonal matrix. 
This problem can in principle be solved by the introduction of 
non-canonical coordinates.\,\cite{lit1} A more heuristic solution,
proposed by Frisk and Guhr,\,\cite{frgu} leads to the determination 
of the dynamics to lowest order in $\hbar$ by the two classical
Hamiltonians
\be
H_{cl}^\pm({\bf r,p}) = \frac{1}{2m}\,p^2+V({\bf r}) 
                         \pm \kappa\,|{\bf B}({\bf r,p})|
\label{lsh}
\ee
whose combined periodic orbits are to be used in the trace formula. 
The first-order $\hbar$ corrections to the energy, in Ref.\,\cite{lit1} 
denoted by $\lambda_{Berry}^{\pm}$ and $\lambda_{NN}^{\pm}$, are to be 
included perturbatively, i.e., only their contributions to the actions 
have to be included in the trace formula. This {\it ad hoc} prescription 
has recently been justified through a relativistic trace formula derived 
from the Dirac equation by Bolte and Keppeler.\,\cite{boke} From the 
non-relativistic reduction of their result, the use of the Hamiltonians 
\eq{lsh} and the above treatment of $\lambda_{Berry}^{\pm}$ and 
$\lambda_{NN}^{\pm}$ can, indeed, be obtained in the ``large-spin'' 
limit.\,\cite{boke}

The equations of motion for the Hamiltonians $H_{cl}^{\pm}$ with 
\eq{Vho} become
\begin{eqnarray}
\dot{r}_i & = & \hspace{0.32cm} p_i \hspace{0.32cm} \pm \; 
                \epsilon_{ijk}\, \vert{\bf B}({\bf r,p})\vert^{-1}
                (B_j\omega_k^2 r_k-B_k\omega_j^2 r_j)\,,\nonumber\\
\dot{p}_i & = & -\omega_i^2 r_i \pm \;
                \epsilon_{ijk}\, \vert{\bf B}({\bf r,p})\vert^{-1}
                (B_j\omega_i^2 p_k - B_k \omega_i^2 p_j)\,. 
                \hspace{0.5cm} i,j,k\in\{x,y,z\} 
\label{eqmot1}
\end{eqnarray}
This is a non-linear system of six equations, and the search for 
periodic orbits is not easy. We have determined them by a 
Newton-Raphson iteration, employing the stability matrix that enters 
the amplitudes in the trace formula.\,\cite{camb} Special care must 
be taken at mode-conversion points where 
${\bf B}({\bf r,p})=0$ and hence the right-hand sides of \eq{eqmot1} 
diverge. Some of the periodic orbits, however, can be found more 
easily. This follows from the fact that the three planes $r_k=p_k=0$ 
$(k=x,y,z)$ in phase space are invariant under the Hamiltonian flow. 
The equations of motion for these two-dimensional orbits are almost 
harmonic:
\begin{eqnarray}
\dot{r}_i & = & \hspace{0.32cm} p_i\hspace{0.32cm}\mp\epsilon_{ijk} 
                \sign (B_k)\,\omega^2_j r_j\,,\nonumber\\
\dot{p}_i & = & -\omega_i^2 r_i \mp \epsilon_{ijk} 
                \sign (B_k)\,\omega_i^2 p_j\,.
                \hspace{1cm} i,j,k \in\{x,y,z\}
\label{eqmot2}
\end{eqnarray} 
Harmonic solutions of \eq{eqmot2} can be obtained analytically by 
ignoring the factors $\sign (B_k)$ that make the equations nonlinear, 
and by accepting only solutions with the correct constant values of 
$\sign (B_k)$. In total, we find six doubly-degenerate planar 
solutions with different frequencies $\omega_{ij}^\pm$ ($i\ne j$)
given by
\begin{equation}
\widetilde{\omega}^\pm_{ij}=\left[
\frac{1}{2}\!\left(\omega_i^2+\omega_j^2+2\kappa^2\omega_i^2\omega_j^2
\pm \sqrt{\left( \omega_i^2-\omega_j^2
\right)^{\!2}\!+8\kappa^2\omega_i^2\omega_j^2
\left(\omega_i^2+\omega_j^2 \right) }\right) \right]^{1/2}\!\!\!\!\!.
\label{omegan}
\end{equation}
These solutions have the form of ellipses, sketched in the left
part of \fig{2dorbits}, which for $\kappa \rightarrow 0$ shrink to 
librations along the $r_i$ axes with the original frequencies 
$\omega_i$. An important fact for these orbits is that $B_k \ne 0$ 
for all $\kappa\ne 0$.
  
It turns out that the frequencies $\widetilde\omega_{ij}^\pm$, 
whose dependence on $\kappa$ is displayed in the right-hand part 
of \fig{2dorbits}, for $\kappa=0.1$ closely agree with the peak 
positions found in \fig{ft}. We cannot resolve all six main peaks; 
a slightly better resolu-

\Figurebb{2dorbits}{85}{543}{531}{693}{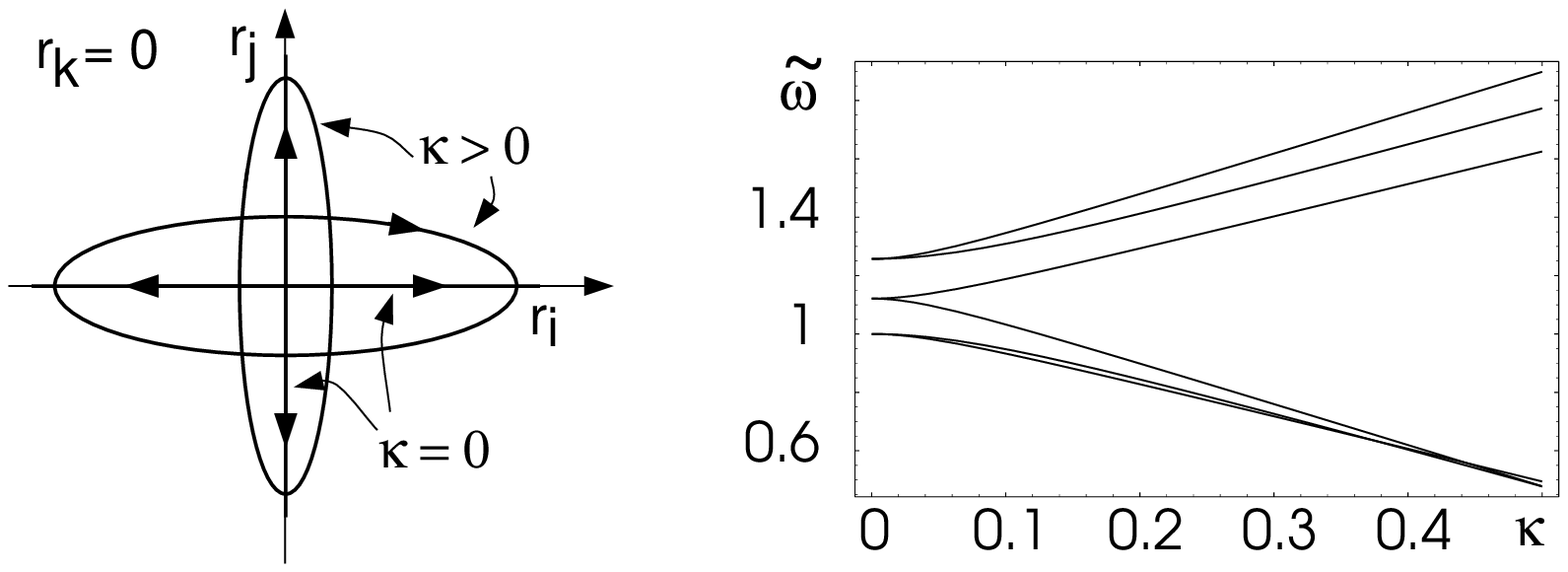}{2.9}{11.8}{
{\it Left:} A schematic plot of a pair of the two-dimensional 
periodic solutions of \protect\eq{eqmot2}. {\it Right:} 
Frequencies $\widetilde{ \omega}_{ij}^\pm$ \protect\eq{omegan} of 
the six planar orbits versus spin-orbit coupling parameter $\kappa$
(using the same parameters $\omega_i$ as in \protect\fig{ft}.)
}

\Figurebb{dge}{15}{125}{766}{535}{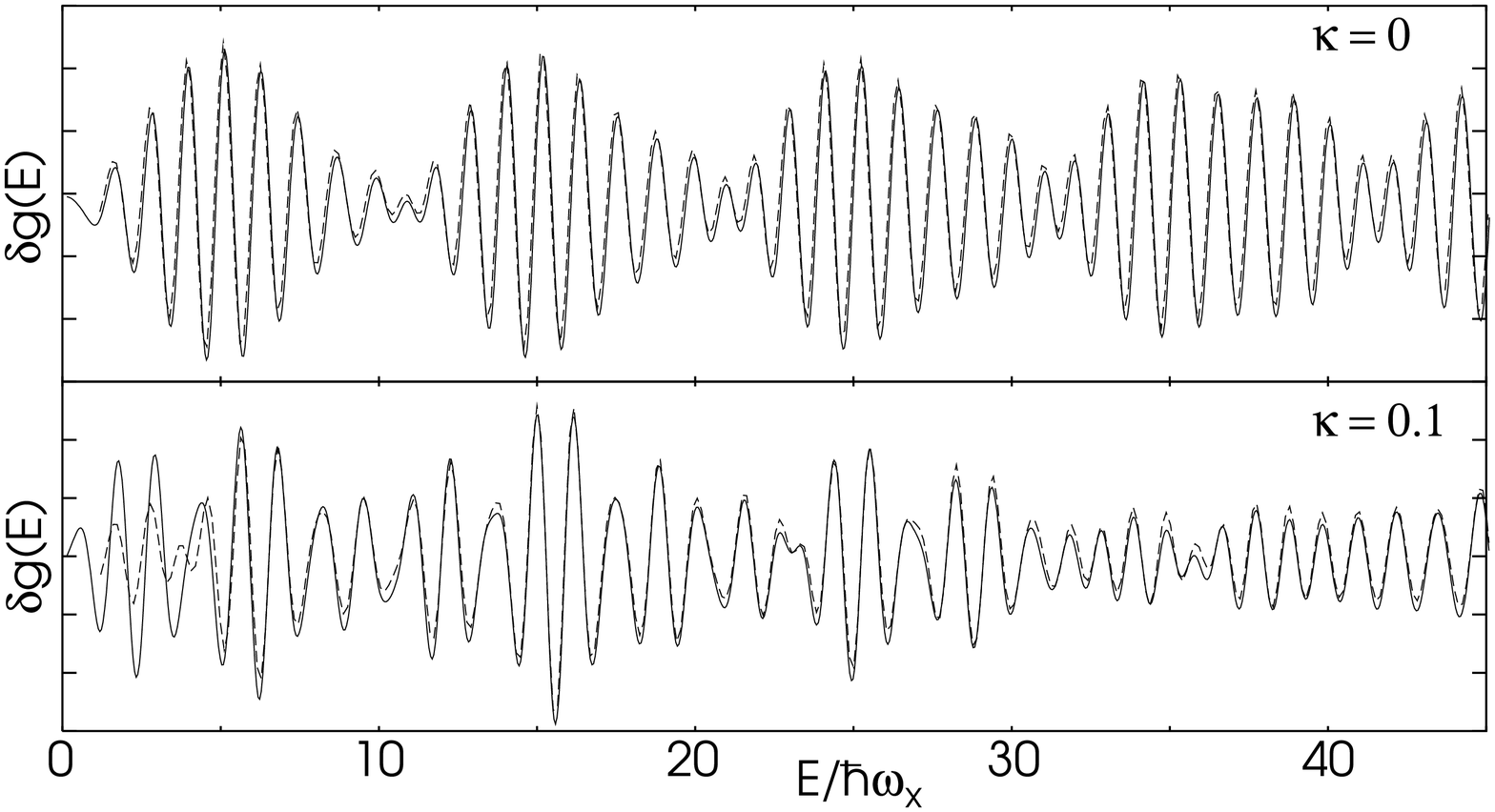}{5.5}{11.8}{
Oscillating part $\delta g\,(E)$ of the level density of the 
3-dimensional anisotropic harmonic oscillator (same parameters as 
in \protect\fig{ft}, but Gaussian averaging range 0.5$\,\hom$). 
{\it Solid lines:} quantum-mechanical results; {\it dotted lines:} 
semiclassical results, for $\kappa=0.1$ using the six planar orbits 
(first repetitions only) in the trace formula.
} 

\noindent
tion is found amongst the peaks corresponding to the second 
harmonics. Some additional small peaks not corresponding to any of 
the $\widetilde\omega_{ij}^\pm$ (e.g., the tiny peak at $T=10.1$) 
can be explained by genuine three-dimensional orbits.\,\cite{camb}
  
We now calculate the oscillating part $\delta g\,(E)$ of the level 
density by including the lowest harmonics of the six planar 
periodic orbits in the trace formula. The $\hbar$-correction 
terms for the planar orbits give a constant shift $\Delta 
S^\pm_{ij}=(\lambda_{Berry}+\lambda_{NN})\,T^\pm_{ij}=\kappa 
\pi\omega_k^2/\widetilde{\omega}^\pm_{ij}$ of the actions at all 
energies.\,\cite{camb} The result for $\delta g\,(E)$ obtained 
with $\kappa=0.1$ is shown in \fig{dge} by the dotted lines. The 
quantum-mechanical results are shown by the solid lines. The 
Gaussian averaging range was here 0.5$\,\hom_x$ in order to 
suppress the contributions from second and higher harmonics. For 
comparison, we also show the curves for $\kappa=0$ which exhibit a 
completely different shell sctructure. A good agreement is found 
also in the presence of the spin-orbit interaction, except for low 
energies where semiclassical approximations usually are worst. 
This shows that the planar orbits are the most important ones for 
the gross-shell structure. The numerical calculation of the 
three-dimensional orbits and their properties is not difficult, as 
long as they are not crossing the mode-conversion surfaces. A 
study of their influence on the finer details of the shell 
structure is in progress.\,\cite{camb} 
  
\section*{Acknowledgment}
This work was supported in parts by the Deutsche 
Forschungsgemeinschaft.

\end{document}